\documentclass[sigconf, fontsize=11pt]{acmart}
\renewcommand\footnotetextcopyrightpermission[1]{} 
\newcommand{\hide}[1]{}

\AtBeginDocument{%
  \providecommand\BibTeX{{%
    \normalfont B\kern-0.5em{\scshape i\kern-0.25em b}\kern-0.8em\TeX}}}

\setcopyright{acmcopyright}
\copyrightyear{2018}
\acmYear{2018}
\acmDOI{XXXXXXX.XXXXXXX}

\acmConference[Conference acronym 'XX]{Make sure to enter the correct
  conference title from your rights confirmation emai}{June 03--05,
  2018}{Woodstock, NY}
\acmPrice{15.00}
\acmISBN{978-1-4503-XXXX-X/18/06}

\begin{document}
\fancyhead[]{}

\title{Privacy-preserving Graph Analytics: \\Secure Generation and Federated Learning}

\author{Dongqi Fu$^\dagger$, Jingrui He$^\dagger$, Hanghang Tong$^\dagger$, and Ross Maciejewski$^\S$}
\affiliation{\institution{$^\dagger$University of Illinois at Urbana-Champaign, $^\S$Arizona State University\\
\{dongqif2, jingrui, htong\}@illinois.edu, rmacieje@asu.edu\\}}

\begin{abstract}
\vspace{-1.75mm}
Directly motivated by security-related applications from the Homeland Security Enterprise, we focus on the privacy-preserving analysis of graph data, which provides the crucial capacity to represent rich attributes and relationships. In particular, we discuss two directions, namely privacy-preserving graph generation and federated graph learning, which can jointly enable the collaboration among multiple parties each possessing private graph data. For each direction, we identify both `quick wins' and `hard problems'. Towards the end, we demonstrate a user interface that can facilitate model explanation, interpretation, and visualization. We believe that the techniques developed in these directions will significantly enhance the capabilities of the Homeland Security Enterprise to tackle and mitigate the various security risks.
\end{abstract}

\maketitle
\vspace{-2.5mm}
\section{Introduction}
\vspace{-1.75mm}
Nowadays, the Homeland Security Enterprise is facing unprecedented challenges in multiple critical areas, such as identifying and preventing targeted violence and mass attacks, building resilient critical infrastructure, countering human trafficking, etc. Addressing these challenges requires collaborative efforts from all levels of government, the private and nonprofit sectors, and individual citizens. In particular, effective and efficient data collection, sharing, analysis, and sense-making are at the core of many decision making processes in these areas. Due to the distributed, sensitive and/or private nature of the large volume of involved data (e.g., personal identifiable information, images and video from surveillance cameras or body camera), it is thus of great importance to make use of the data while avoiding the sharing and use of sensitive information. In this paper, we focus on graph data, or network data, due to their rich representation capabilities to encode the multi-modality time-evolving attributes for entities (e.g., individuals, locations) as node attributes in the graph, and to encode the various types of relationships among entities via edges.

\vspace{-0.45mm}

The main goal of this paper is to focus on privacy-preserving analysis of graphs. In particular, we aim to explore key research questions and privacy-enhancing technologies that target the following two areas: (A1) privacy-preserving techniques for creating, maintaining and linking anonymous identities and profiles; and (A2) multiparty and homomorphic computation to allow for analysis of datasets held by multiple parties without the need to physically combine datasets. The techniques developed for these areas together will enable the collaboration among multiple parties each possessing private graph data.

\vspace{-0.45mm}

More specifically, we consider the following two directions for the two areas respectively. For (A1), a promising direction is to generate synthetic graphs in a privacy-preserving manner, such that the generated graphs can be shared with collaborators without revealing sensitive information on either the node level (node-DP) or the edge level (edge-DP). For (A2), a promising direction is to perform federated learning among multiple parties, or clients, such that the central server can build robust, effective and efficient predictive models while preserving the privacy of individual clients. Next, we elaborate on the two directions, including both `quick wins' and `hard problems', followed by additional discussions regarding the development of the user interface that could enable subject matter experts to make effective use of the developed techniques.

\vspace{-3mm}

\section{Privacy-preserving Generation}
\vspace{-1.5mm}

Graphs represent complex relational information between entities, such that modeling the graph generation process and then generating many more meaningful graphs could contribute to various applications~\cite{DBLP:journals/csur/BonifatiHPS20}.
However, mimicking the observed graph as much as possible will induce a privacy risk after the generation~\cite{DBLP:series/ads/WuYLC10}. For example, a node's identity is highly likely to be exposed in the generated social network if its connections are mostly preserved, which means a degree-based node attacker will easily detect a vulnerability in the generated graph with some background knowledge.

\vspace{-3mm}

\subsection{Quick Wins}
\vspace{-1.5mm}
For privacy-preserving static graph generation, current solutions can be roughly classified into two types that can be readily applied.
First, the anonymization is directly performed on the observed topology to generate new graph data, such as randomizing the adjacency~\cite{DBLP:conf/sdm/YangW08}, injecting the connection uncertainty~\cite{DBLP:conf/ccs/NguyenIR15}, or permutating the connection distribution under the edge-level differential privacy (edge-DP)~\cite{DBLP:conf/ccs/QinYYKX017}.
Second, following the synergy of deep learning and differential privacy~\cite{DBLP:conf/ccs/AbadiCGMMT016}, deep generative models for graphs are recently proposed under privacy constraints. To be specific, in~\cite{DBLP:conf/ijcai/YangWZCS21}, the privacy scheme is added to the gradient descent phase of the generation learning process.

\vspace{-3mm}

\subsection{Hard Problems}
\vspace{-1.5mm}
Most, if not all, of privacy-preserving graph generation methods consider the observed graphs as static. However, in the complex real-world scenarios, the graphs are usually evolving over time~\cite{DBLP:conf/kdd/FuZH20, DBLP:conf/cikm/FuXLTH20, DBLP:conf/sigir/FuH21, DBLP:journals/corr/abs-2107-02168, DBLP:journals/corr/abs-2203-04928}, which brings critical challenges to the current privacy-preserving static graph generation process. \hide{In other words, the time domain enriches the node attribute dimension and may also dictate the attribute distribution, which leads to increasing exposure risk.} To the best of our knowledge, how to generate privacy-preserving temporal graphs largely remains open. For example, (1) unlike the abundant research on static graphs, what kind of time-aware information is sensitive and should be hidden in the generated graph to protect entities' privacy is not clear; (2) even if the sensitive information is determined, the time-aware protection mechanism is not yet available; (3) once the protection mechanism is designed, it can be challenging to maintain the generation utility at the same time with privacy constraints.

\vspace{-3mm}

\section{Federated Learning with Graphs}\vspace{-1.5mm}
According to~\cite{49232}, "Federated learning (FL) is a machine learning setting where many clients (e.g. mobile devices or whole organizations) collaboratively train a model under the orchestration of a central server (e.g. service provider), while keeping the training data decentralized." This problem setting is particularly important for privacy-preserving analysis of private graph data that might reside on multiple clients such as the servers for various organizations. In other words, it enables centralized decision making using all available graph data, while avoiding physically combining datasets.

In the past few years, federated learning has been extensively studied, focusing on research problems such as model robustness to adversarial attacks, and data distribution among clients. Despite some existing work using graph data (e.g.,~\cite{abs-2104-07145}), robust federated learning with graph data largely remains under-explored.

\vspace{-3mm}

\subsection{Quick Wins}\vspace{-1.25mm}
In general, federated learning systems can be vulnerable to attacks and failures, e.g., Byzantine attacks can lead to the convergence to an unsatisfactory model or even divergence~\cite{krum}. A common defending mechanism is to replace the gradients
averaging with robust estimation of the center~\cite{krum}. These methods have proven Byzantine-robustness when data from different clients are independent and identically distributed (IID). On one hand, these methods can be directly adapted to model graph data by using, e.g., graph neural networks~\cite{abs-1901-00596}. On the other hand, the IID assumption regarding client data can fail miserably due to the special properties of graph data (e.g., homophily) as compared to other types of data. 

To address this problem, one `quick win' would be to observe the performance of existing Byzantine-robust methods on graph data, in order to study the impact of IID violation to the performance of the centralized model. Based on the results from this study, another `quick win' would be to develop robust federated learning methods tailored for graph data that violate the IID assumption. For example, in our previous work, we studied a special type of non-IIDness among client data, i.e., label skewness, and proposed a two-stage algorithm to estimate the true parameters of the centralized model in the presence of Byzantine clients. Based on this work, one can further study other types of non-IIDness among client data that may fit various types of graph data, and design the robust algorithms.

\vspace{-3mm}

\subsection{Hard Problems}\vspace{-1.5mm}
For privacy-preserving analysis of graph data, one major benefit of federated learning is that it could avoid physically combining private data from individual clients. On the other hand, some recent studies on federated learning require the availability of clean and non-sensitive data on the server (e.g., to enable knowledge distillation in the presence of heterogeneous models among clients~\cite{LinKSJ20}), which can potentially be shared with clients without violating the privacy constraints. However, such data can be difficult to obtain, especially in security related applications (e.g., the identification of targeted violence). Despite some recent work focusing on data-free knowledge distillation for federated learning~\cite{ZhuHZ21}, the proposed solutions cannot be readily applied on graph data due to the significant difference in generating graph data vs. non-graph data. To solve this `hard problem', the graph generative models (as discussed in the previous section) for both static and dynamic graphs can potentially be integrated into the federated learning pipeline, to enable the data-free robust federated learning for graph data.

\vspace{-3mm}
\section{User Interface}\vspace{-1.5mm}
With the development of novel privacy-enhancing technologies, one critical aspect is model explanation, interpretation, and visualization facing various audience. This calls for effective and efficient user interfaces that serve three major purposes: (1) to solicit feedback from subject matter experts for model improvement; (2) to empower subject matter experts with deep insights regarding the model and the data; (3) to aid decision making in various security related applications. \hide{Furthermore, the user interface should be tailored for the specific audience, e.g., the central server vs. individual clients in federated learning, as they have access to different amount of information and may have varying design objectives.} The following figure exemplifies such an interface that we developed for explaining transfer learning~\cite{MaFHNM21}.

\vspace{-3mm}

\begin{figure}[H]
    \centering
    \includegraphics[width=1\linewidth]{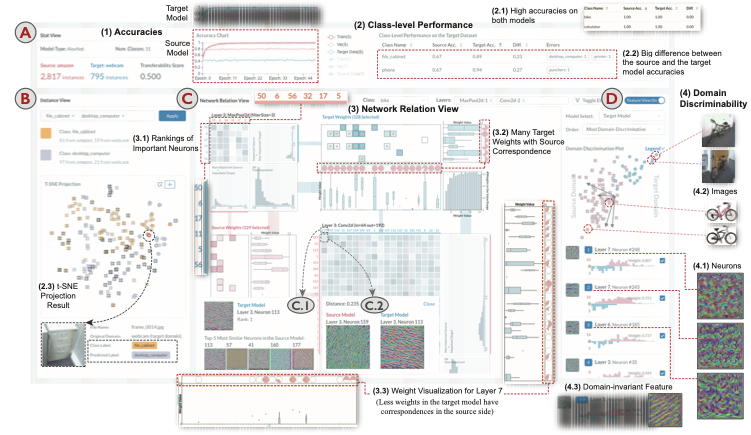}\vspace{-4mm}
    \caption{Visual Analytics for Transfer Learning Interface}
    \label{fig:tl_explain}
\end{figure}

\vspace{-6mm}

\section{Conclusion}\vspace{-1mm}
In this paper, we discussed privacy-preserving analysis of graph data in two directions, namely secure generation and federated learning. We believe that research efforts dedicated to these directions can lead to significantly enhanced capabilities of the Homeland Security Enterprise for countering the various security risks.



\vspace{-1.4mm}
\small
\bibliographystyle{ACM-Reference-Format}
\bibliography{ref.bib}


\begin{thebibliography}{19}


\ifx \showCODEN    \undefined \def \showCODEN     #1{\unskip}     \fi
\ifx \showDOI      \undefined \def \showDOI       #1{#1}\fi
\ifx \showISBNx    \undefined \def \showISBNx     #1{\unskip}     \fi
\ifx \showISBNxiii \undefined \def \showISBNxiii  #1{\unskip}     \fi
\ifx \showISSN     \undefined \def \showISSN      #1{\unskip}     \fi
\ifx \showLCCN     \undefined \def \showLCCN      #1{\unskip}     \fi
\ifx \shownote     \undefined \def \shownote      #1{#1}          \fi
\ifx \showarticletitle \undefined \def \showarticletitle #1{#1}   \fi
\ifx \showURL      \undefined \def \showURL       {\relax}        \fi
\providecommand\bibfield[2]{#2}
\providecommand\bibinfo[2]{#2}
\providecommand\natexlab[1]{#1}
\providecommand\showeprint[2][]{arXiv:#2}

\bibitem[Abadi et~al\mbox{.}(2016)]%
        {DBLP:conf/ccs/AbadiCGMMT016}
\bibfield{author}{\bibinfo{person}{M. Abadi}, \bibinfo{person}{A. Chu},
  \bibinfo{person}{I.~J. Goodfellow}, \bibinfo{person}{H.~B. McMahan},
  \bibinfo{person}{I. Mironov}, \bibinfo{person}{K. Talwar}, {and}
  \bibinfo{person}{L. Zhang}.} \bibinfo{year}{2016}\natexlab{}.
\newblock \showarticletitle{Deep Learning with Differential Privacy}. In
  \bibinfo{booktitle}{\emph{CCS 2016}}.
\newblock


\bibitem[Blanchard et~al\mbox{.}(2017)]%
        {krum}
\bibfield{author}{\bibinfo{person}{P. Blanchard}, \bibinfo{person}{E.~M.~El
  Mhamdi}, \bibinfo{person}{R. Guerraoui}, {and} \bibinfo{person}{J. Stainer}.}
  \bibinfo{year}{2017}\natexlab{}.
\newblock \showarticletitle{Machine Learning with Adversaries: Byzantine
  Tolerant Gradient Descent}. In \bibinfo{booktitle}{\emph{NeurIPS 2017}}.
\newblock


\bibitem[Bonifati et~al\mbox{.}(2020)]%
        {DBLP:journals/csur/BonifatiHPS20}
\bibfield{author}{\bibinfo{person}{A. Bonifati}, \bibinfo{person}{I.
  Holubov{\'{a}}}, \bibinfo{person}{A. Prat{-}P{\'{e}}rez}, {and}
  \bibinfo{person}{S. Sakr}.} \bibinfo{year}{2020}\natexlab{}.
\newblock \showarticletitle{Graph Generators: State of the Art and Open
  Challenges}.
\newblock \bibinfo{journal}{\emph{{ACM} Comput. Surv.}} (\bibinfo{year}{2020}).
\newblock


\bibitem[Fu et~al\mbox{.}(2022)]%
        {DBLP:journals/corr/abs-2203-04928}
\bibfield{author}{\bibinfo{person}{D. Fu}, \bibinfo{person}{Y. Ban},
  \bibinfo{person}{H. Tong}, \bibinfo{person}{R. Maciejewski}, {and}
  \bibinfo{person}{J. He}.} \bibinfo{year}{2022}\natexlab{}.
\newblock \showarticletitle{{DISCO:} Comprehensive and Explainable
  Disinformation Detection}.
\newblock \bibinfo{journal}{\emph{CoRR}} (\bibinfo{year}{2022}).
\newblock


\bibitem[Fu and He(2021a)]%
        {DBLP:journals/corr/abs-2107-02168}
\bibfield{author}{\bibinfo{person}{D. Fu} {and} \bibinfo{person}{J. He}.}
  \bibinfo{year}{2021}\natexlab{a}.
\newblock \showarticletitle{{DPPIN:} {A} Biological Repository of Dynamic
  Protein-Protein Interaction Network Data}.
\newblock \bibinfo{journal}{\emph{CoRR}} (\bibinfo{year}{2021}).
\newblock


\bibitem[Fu and He(2021b)]%
        {DBLP:conf/sigir/FuH21}
\bibfield{author}{\bibinfo{person}{D. Fu} {and} \bibinfo{person}{J. He}.}
  \bibinfo{year}{2021}\natexlab{b}.
\newblock \showarticletitle{{SDG:} {A} Simplified and Dynamic Graph Neural
  Network}. In \bibinfo{booktitle}{\emph{SIGIR 2021}}.
\newblock


\bibitem[Fu et~al\mbox{.}(2020a)]%
        {DBLP:conf/cikm/FuXLTH20}
\bibfield{author}{\bibinfo{person}{D. Fu}, \bibinfo{person}{Z. Xu},
  \bibinfo{person}{B. Li}, \bibinfo{person}{H. Tong}, {and} \bibinfo{person}{J.
  He}.} \bibinfo{year}{2020}\natexlab{a}.
\newblock \showarticletitle{A View-Adversarial Framework for Multi-View Network
  Embedding}. In \bibinfo{booktitle}{\emph{CIKM 2020}}.
\newblock


\bibitem[Fu et~al\mbox{.}(2020b)]%
        {DBLP:conf/kdd/FuZH20}
\bibfield{author}{\bibinfo{person}{D. Fu}, \bibinfo{person}{D. Zhou}, {and}
  \bibinfo{person}{J. He}.} \bibinfo{year}{2020}\natexlab{b}.
\newblock \showarticletitle{Local Motif Clustering on Time-Evolving Graphs}. In
  \bibinfo{booktitle}{\emph{KDD 2020}}.
\newblock


\bibitem[He et~al\mbox{.}(2021)]%
        {abs-2104-07145}
\bibfield{author}{\bibinfo{person}{C. He}, \bibinfo{person}{K.
  Balasubramanian}, \bibinfo{person}{E. Ceyani}, \bibinfo{person}{Y. Rong},
  \bibinfo{person}{P. Zhao}, \bibinfo{person}{J. Huang}, \bibinfo{person}{M.
  Annavaram}, {and} \bibinfo{person}{S. Avestimehr}.}
  \bibinfo{year}{2021}\natexlab{}.
\newblock \showarticletitle{FedGraphNN: {A} Federated Learning System and
  Benchmark for Graph Neural Networks}.
\newblock \bibinfo{journal}{\emph{CoRR}} (\bibinfo{year}{2021}).
\newblock


\bibitem[Kairouz and et~al.(2019)]%
        {49232}
\bibfield{author}{\bibinfo{person}{P. Kairouz} {and} \bibinfo{person}{et al.}}
  \bibinfo{year}{2019}\natexlab{}.
\newblock \showarticletitle{Advances and Open Problems in Federated Learning}.
\newblock


\bibitem[Lin et~al\mbox{.}(2020)]%
        {LinKSJ20}
\bibfield{author}{\bibinfo{person}{T. Lin}, \bibinfo{person}{L. Kong},
  \bibinfo{person}{S.~U. Stich}, {and} \bibinfo{person}{M. Jaggi}.}
  \bibinfo{year}{2020}\natexlab{}.
\newblock \showarticletitle{Ensemble Distillation for Robust Model Fusion in
  Federated Learning}. In \bibinfo{booktitle}{\emph{NeurIPS 2020}}.
\newblock


\bibitem[Ma et~al\mbox{.}(2021)]%
        {MaFHNM21}
\bibfield{author}{\bibinfo{person}{Y. Ma}, \bibinfo{person}{A. Fan},
  \bibinfo{person}{J. He}, \bibinfo{person}{A.~Reddy Nelakurthi}, {and}
  \bibinfo{person}{R. Maciejewski}.} \bibinfo{year}{2021}\natexlab{}.
\newblock \showarticletitle{A Visual Analytics Framework for Explaining and
  Diagnosing Transfer Learning Processes}.
\newblock \bibinfo{journal}{\emph{{IEEE} Trans. Vis. Comput. Graph.}}
  (\bibinfo{year}{2021}).
\newblock


\bibitem[Nguyen et~al\mbox{.}(2015)]%
        {DBLP:conf/ccs/NguyenIR15}
\bibfield{author}{\bibinfo{person}{H.~H. Nguyen}, \bibinfo{person}{A. Imine},
  {and} \bibinfo{person}{M. Rusinowitch}.} \bibinfo{year}{2015}\natexlab{}.
\newblock \showarticletitle{Anonymizing Social Graphs via Uncertainty
  Semantics}. In \bibinfo{booktitle}{\emph{CCS 2015}}.
\newblock


\bibitem[Qin et~al\mbox{.}(2017)]%
        {DBLP:conf/ccs/QinYYKX017}
\bibfield{author}{\bibinfo{person}{Z. Qin}, \bibinfo{person}{T. Yu},
  \bibinfo{person}{Y. Yang}, \bibinfo{person}{I. Khalil}, \bibinfo{person}{X.
  Xiao}, {and} \bibinfo{person}{K. Ren}.} \bibinfo{year}{2017}\natexlab{}.
\newblock \showarticletitle{Generating Synthetic Decentralized Social Graphs
  with Local Differential Privacy}. In \bibinfo{booktitle}{\emph{CCS 2017}}.
\newblock


\bibitem[Wu et~al\mbox{.}(2010)]%
        {DBLP:series/ads/WuYLC10}
\bibfield{author}{\bibinfo{person}{X. Wu}, \bibinfo{person}{X. Ying},
  \bibinfo{person}{K. Liu}, {and} \bibinfo{person}{L. Chen}.}
  \bibinfo{year}{2010}\natexlab{}.
\newblock \showarticletitle{A Survey of Privacy-Preservation of Graphs and
  Social Networks}.
\newblock In \bibinfo{booktitle}{\emph{Managing and Mining Graph Data}}.
\newblock


\bibitem[Wu et~al\mbox{.}(2019)]%
        {abs-1901-00596}
\bibfield{author}{\bibinfo{person}{Z. Wu}, \bibinfo{person}{S. Pan},
  \bibinfo{person}{F. Chen}, \bibinfo{person}{G. Long}, \bibinfo{person}{C.
  Zhang}, {and} \bibinfo{person}{P.~S. Yu}.} \bibinfo{year}{2019}\natexlab{}.
\newblock \showarticletitle{A Comprehensive Survey on Graph Neural Networks}.
\newblock \bibinfo{journal}{\emph{CoRR}} (\bibinfo{year}{2019}).
\newblock


\bibitem[Yang et~al\mbox{.}(2021)]%
        {DBLP:conf/ijcai/YangWZCS21}
\bibfield{author}{\bibinfo{person}{C. Yang}, \bibinfo{person}{H. Wang},
  \bibinfo{person}{K. Zhang}, \bibinfo{person}{L. Chen}, {and}
  \bibinfo{person}{L. Sun}.} \bibinfo{year}{2021}\natexlab{}.
\newblock \showarticletitle{Secure Deep Graph Generation with Link Differential
  Privacy}. In \bibinfo{booktitle}{\emph{IJCAI 2021}}.
\newblock


\bibitem[Ying and Wu(2008)]%
        {DBLP:conf/sdm/YangW08}
\bibfield{author}{\bibinfo{person}{X. Ying} {and} \bibinfo{person}{X. Wu}.}
  \bibinfo{year}{2008}\natexlab{}.
\newblock \showarticletitle{Randomizing Social Networks: a Spectrum Preserving
  Approach}. In \bibinfo{booktitle}{\emph{SDM 2008}}.
\newblock


\bibitem[Zhu et~al\mbox{.}(2021)]%
        {ZhuHZ21}
\bibfield{author}{\bibinfo{person}{Z. Zhu}, \bibinfo{person}{J. Hong}, {and}
  \bibinfo{person}{J. Zhou}.} \bibinfo{year}{2021}\natexlab{}.
\newblock \showarticletitle{Data-Free Knowledge Distillation for Heterogeneous
  Federated Learning}. In \bibinfo{booktitle}{\emph{ICML 2021}}.
\newblock


\end{thebibliography}

\end{document}